\documentclass[trackchanges]{aastex701}

\begin{document}

\title{Compact Absorbers surrounding the M87 AGN Revealed by ALMA Band 3 Observations}

\author[orcid=0000-0001-7482-4967, gname=Mahitosh, sname=Ray]{Mahitosh Ray}
\affiliation{Graduate Institute of Astronomy, National Central University, 300 Zhongda Road, Taoyuan, Taiwan}
\email[show]{d1109601@gm.astro.ncu.edu.tw}
\author[gname=Chorng-Yuan, sname=Hwang]{Chorng-Yuan Hwang}
\affiliation{Graduate Institute of Astronomy, National Central University, 300 Zhongda Road, Taoyuan, Taiwan}
\email[show]{hwangcy@astro.ncu.edu.tw}
\begin{abstract}

We report detection of two compact absorption features surrounding the M87 nucleus and some extended absorption features perpendicular to the M87 jet using the Atacama Large millimeter-submillimeter-Array (ALMA) Band 3 data. One compact absorption feature appears at the position of the M87 AGN and the other $0.5\arcsec$ away from the AGN. These two compact features appear as  well-defined negative structures in the integrated intensity map from the velocity range of $-500$ to $+2000~{\rm km~s^{-1}}$. These two features are separated by a distance of $\sim$40~pc assuming the distance of M87. The origin of these features could be dense molecular fragments associated with super massive black holes (SMBHs), possibly associated with a rotating torus filament, or in-falling molecular fragments/objects feeding the nucleus. The extended absorption features perpendicular to the jet appear in all velocity ranges and can be caused by shock compressed regions associated with the jet knots A and C. 

\end{abstract}


\section{Introduction}\label{sec:intro}
At the hearts of many galaxy clusters lie cD galaxies; the most massive and luminous galaxies known to often host colossal super massive black holes (SMBHs) with masses exceeding $10^9~\mathrm{M_\odot}$. These gravitational behemoths play a vital role in galactic dynamics, feedback processes, and the regulation of star formation in their host galaxies \citep{Cattaneo...}. The SMBHs in cD galaxies frequently manifest as active galactic nuclei (AGNs), which emit vast amounts of energy across the electromagnetic spectrum as they accrete the surrounding gas \citep{shin...}. AGN activity is not merely a local phenomenon. The activity is frequently associated with the production of powerful relativistic jets, which extend well beyond the host galaxy. These high velocity jets could create shocks in the interstellar and intergalactic medium and could serve as both diagnostics of SMBH activity and agents of feedback, capable of heating and displacing gas on galactic scales \citep{gutierrez...}. 

Recent studies suggest that cD galaxies, owing to their complex merger histories, 
may host dual or merging supermassive black hole (SMBH) systems \citep{Lupi...}. 
Such SMBH mergers are expected to produce distinctive gravitational and 
electromagnetic signatures, particularly when embedded in dense, gas-rich 
environments such as AGN accretion disks \citep{Accretion..., Tagawa..}. 
The presence of a secondary SMBH could induce measurable dynamical effects on the primary SMBH and, consequently, on the jet launched from its vicinity. In particular, a secondary SMBH may cause a displacement of the primary SMBH from the galaxy’s center of mass \citep{SMBH_offset...}, as well as reflex motion that can manifest as jet precession, transverse oscillations, or other quasi-periodic structural variations \citep{secondary_BH...}.  Such jet perturbations provide an indirect but 
powerful observational probe of close-separation SMBH binaries, even when 
the secondary black hole itself remains unresolved. Direct observational evidence for SMBH mergers remains scarce; however, a few notable examples of galaxies hosting merging or multiple SMBH systems have been 
reported in the literature \citep{merger_ngc6240...,triple_galactic_merger...}.

The M87 galaxy is a massive cD galaxy at the center of the Virgo cluster. This cD galaxy houses a SMBH with a mass of approximately $\mathrm{6.5\times10^9~M_\odot}$ \citep{EHT...} and encapsulates many of the AGN activities. This SMBH powers one of the best studied relativistic jets
observed in many different wave bands extending over several kilo-parsecs due to its proximity,
exceptional brightness, and the availability of high-resolution, multi-wavelength observations \citep{Biretta..., VLA_abc..., radio_to_Xray...}. Previous observations of periodic features in the M87 jet suggested a possible secondary SMBH separated from the primary one by $0.01- 0.05$ pc \citep{secondary_BH...}. Besides, a projected displacement of the nuclear point source in M87 by $\sim7$ pc from the galactic photometric center towards the counter-jet direction, aligned with the jet axis, has been interpreted as possible evidence for dynamical perturbations of the SMBH due to massive objects \citep{SMBH_offset...}. However, subsequent dynamical studies have found no compelling evidence for a long-lived offset, favoring a single dominant central SMBH in M87 \citep{walsh2013...}.

In this paper, we present the observational results of the central region of the M87 in the vicinity of the SMBH using the Atacama Large millimeter sub-millimeter Array (ALMA) archival data of Band 3 observation of the CO(1-0) rotational transition.   
 
\section{Data reduction and analysis}\label{sec:Data reduction and analysis}

In this paper, we use the ALMA archival data for CO(1-0) rotational transition from the ALMA science portal at the National Radio Astronomy Observatory (NRAO) under the project code 2016.1.00021.S. 
The data were calibrated and analyzed using the Common Astronomy Software Application (CASA) software package (version 5.6.1), Cube Analysis and Rendering Tool for Astronomy (CARTA version 4.0.0). The calibrated data for the CO line was reprocessed based on the scripts provided by the archival data. We carried out continuum subtractions and created channel maps according to our requirements.

In our analysis, we used all the available spectral windows (SPWs) to create the continuum map. For cube imaging we only used SPW 29 because it has much better signal to noise ratio (SNR) than the other 2 available windows. The data were converted to image cubes using \texttt{specmode=cube} in the \texttt{TCLEAN} task of CASA. We adopt Hogbom deconvolver with the Briggs robust weighting of 0.5 and image size of 2500 $\times$ 2500 pixels with the cell size of 0.035 arc seconds (estimated using the analysis utilities task of CASA). 

We generated cleaned image cubes using \texttt{niter=1000} and utilized the non-interactive (auto clean) method. The resulting synthesized beam size is $0.22\arcsec \times 0.21\arcsec$ . To perform continuum subtraction, we used the \texttt{UVCONTSUB} task with \texttt{fitorder=1} and selected line-free channels (Figure 1 shows the continuum emission image). The spectrum shown in Figure 2 is obtained from an elliptical regions slightly larger than the beam size of the images ($0.26\arcsec \times 0.26\arcsec$) centered at M87 (R.A.: $12^h30^m49^s.423$, Decl.: $12^h23^m28^s.043$). 

\begin{figure}[ht]
    \centering
    \includegraphics[width=12cm, height = 8cm]{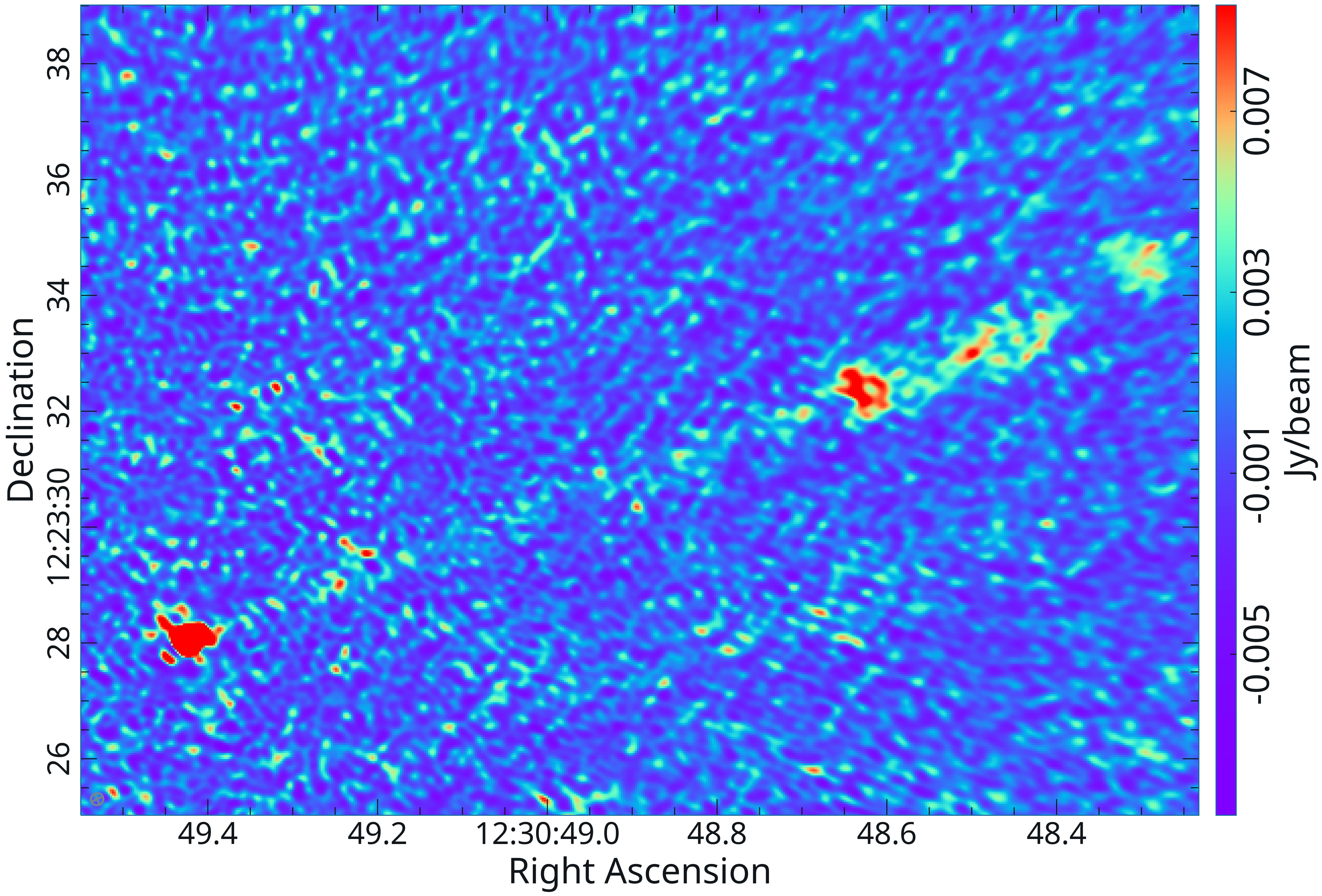}
    \caption{Continuum emission image @114.04 GHz of nucleus and jet of M87.}
\label{fig:M87 continuum emission image}    
\end{figure}

Figure 2 shows the CO(1-0) continuum subtracted spectra of the central region of the M87 galaxy, R.A.:$\mathrm{12^h30^m49^s.423}$ Dec.:$\mathrm{12^h23^m28^s.043}$. with a selected region size of $\mathrm{0.6\arcsec \times 0.6\arcsec}$. The spectrum shows a broad absorption profile starting from $\sim$ 200 to 2000 $\mathrm{km~s^{-1}}$ with a systemic velocity of 1266$\mathrm{km~s^{-1}}$. The broad absorption profile is closely similar to the results shown in \citet{Ray...}, which have much lower spectral and spatial resolutions. This suggests that this absorption feature should be real and is associated with the central AGN.

\begin{figure*}[ht!]
\centering
\includegraphics[width=11cm, height=7cm]{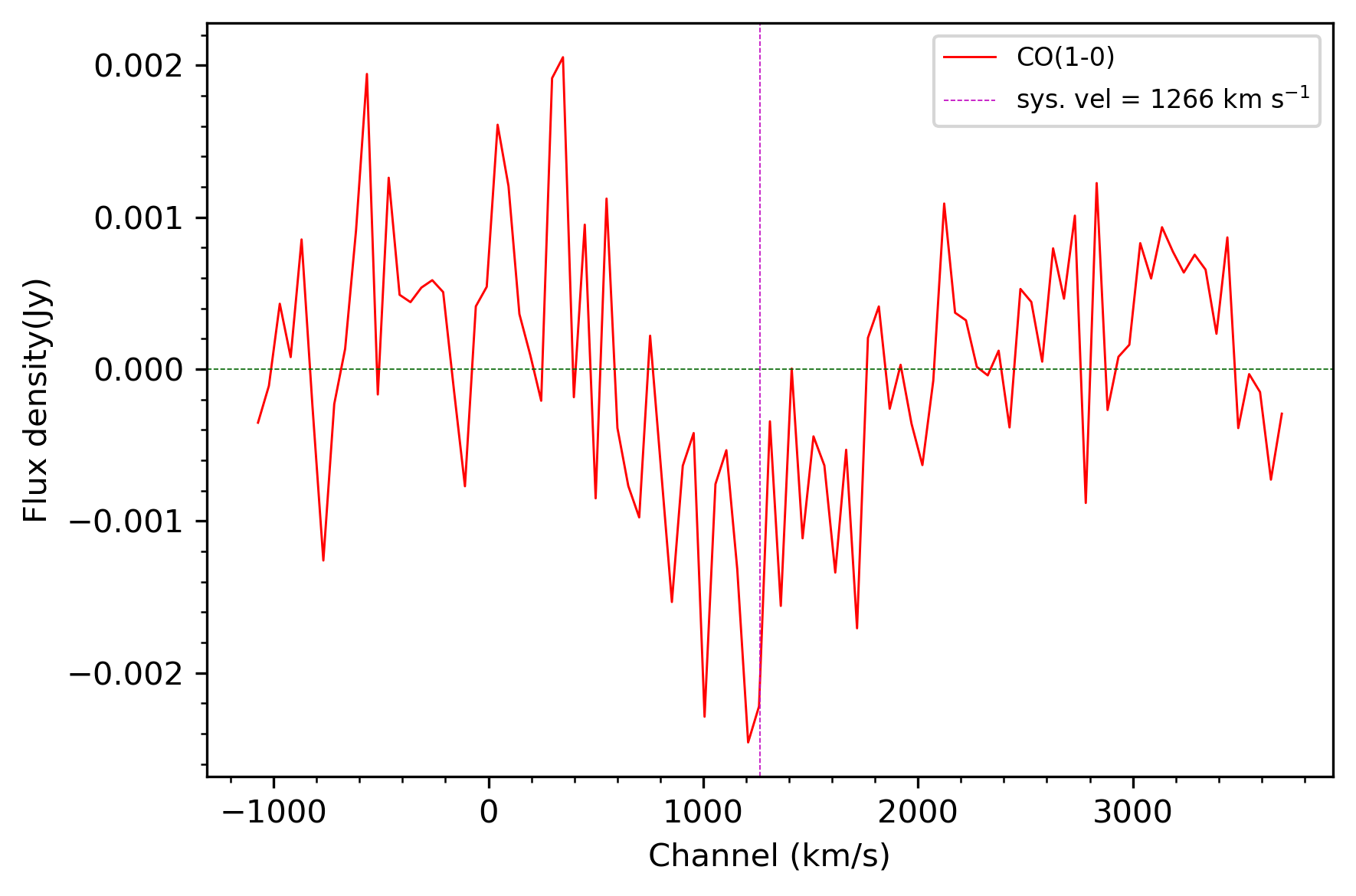}
\caption{Observed spectrum of the M87 central region around the SMBH.)}.
\label{fig:M87 spectrum}
\end{figure*}

\section{Results and Discussion}\label{sec:results and discussion}
\subsection{Absorption features}
\label{subsec:observed features}
To show the absorption distribution, we created a moment 0 map using the pixel values of the channel image $\le 0$. Here we used $\le 0$ because we want to create absorption distribution, i.e., the negative part of the spectrum. We integrated over the velocity range $500$ to $2000~{\rm km~s^{-1}}$ to obtain the moment 0 map and detected two compact negative features in the image (Figure 3).  
The observed features appear to be two spatially localized dark spots close to the line of sight to the AGN. One spot appears along the line of sight to the AGN and we identify it as ALMA-1. The other spot appears to be less absorbed and is $0.5\arcsec$ away from ALMA-1; we identify the less absorbed spot as ALMA-2. The two compact structures are separated by a distance of $\sim 40$~pc measured from the center of ALMA-1 to the center of ALMA-2 assuming the distance of M87. Another thing to note is that ALMA-2 is in the general opposite direction of the M87 jet.

\begin{figure*}[ht!]
\centering
\includegraphics[width=8cm, height=7cm]{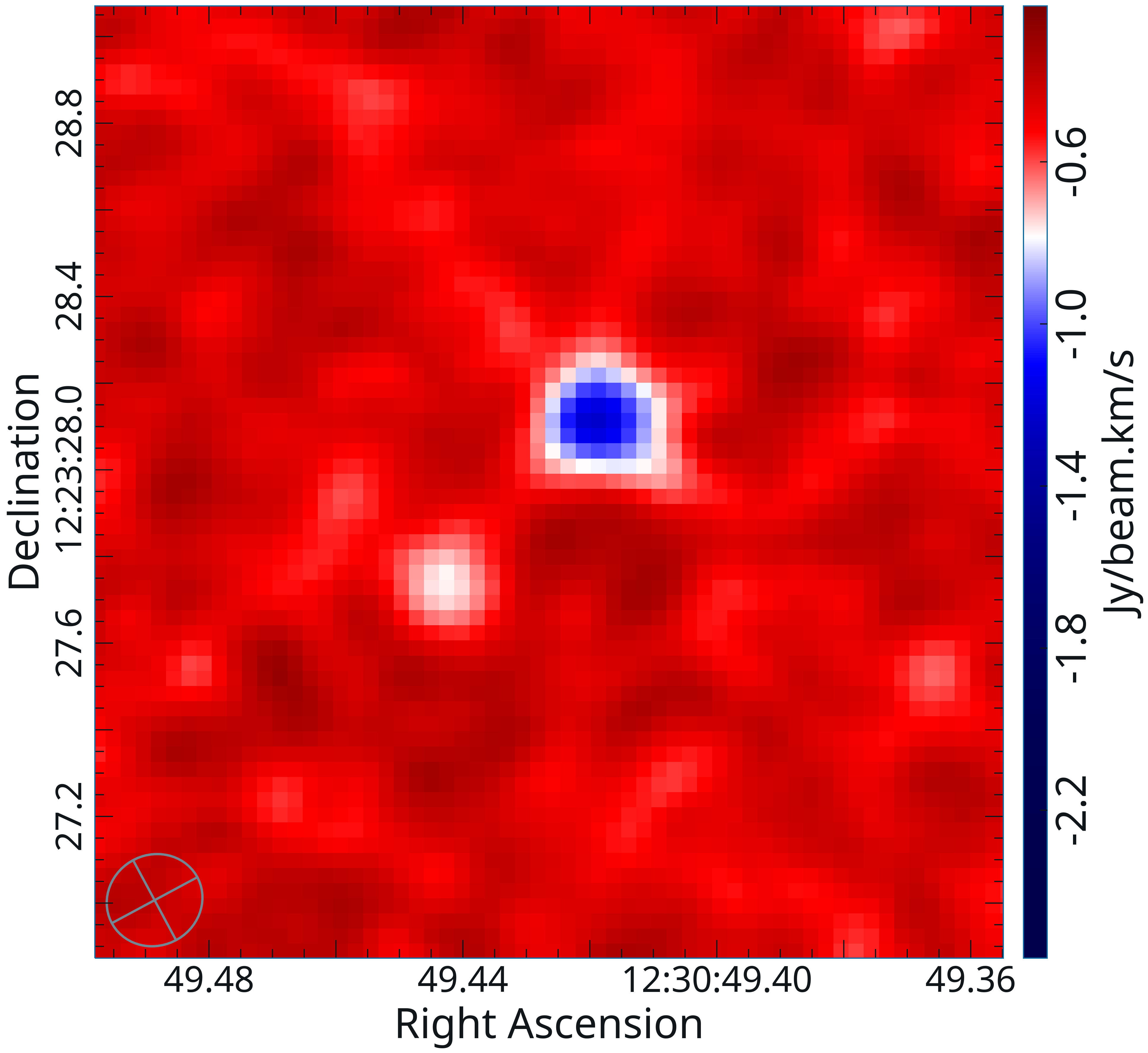}
\includegraphics[width=8cm, height=7cm]{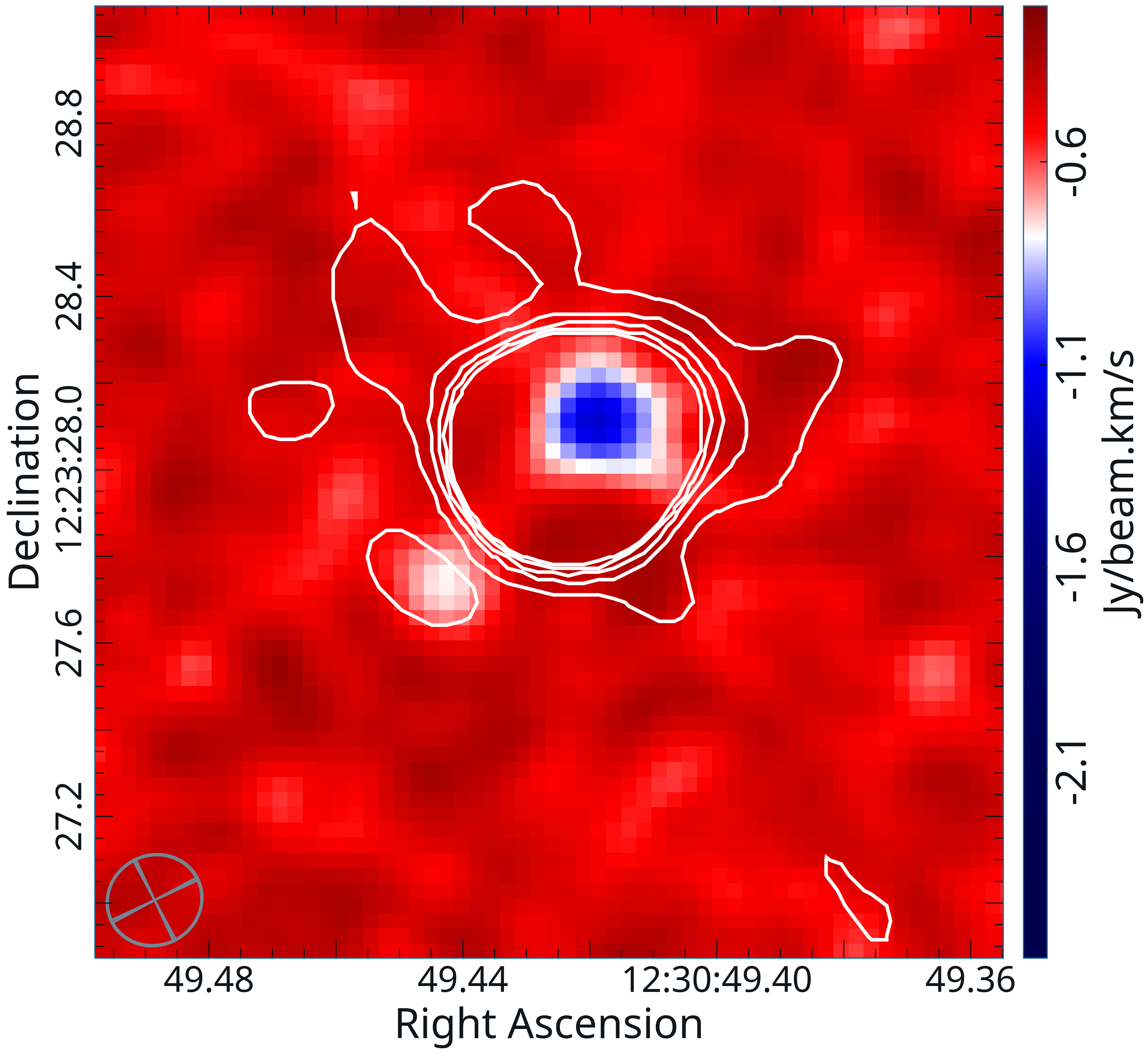}
\caption{Left: Observed detection of the 2 absorption component in the central region of M87. The strong absorber is on the right and is in front of the AGN (ALMA-1), the weak absorber is the one on the left (ALMA-2). Right: Observed absorption components overlapped with the continuum emission (white contours) at 1, 3, 5, 7 and 9 $\sigma$.}
\label{fig:ALMA-1 and ALMA-2}
\end{figure*}

Both ALMA-1 and ALMA-2 appear only in the above mentioned velocity range. ALMA-2 will disappear if the moment 0 map is created outside of this velocity range. When the map is created by integrating all channels in the cube, the structure becomes diluted and is less prominent. 

Another important features are the line-shaped absorption features perpendicular to the M87 jet shown in Figure 4. We observed some line-shaped absorbed regions in the integrated absorption map around knots A, B, and C, which is identified in the Very Large array (VLA) data \citep{VLA_abc...}. These line-shaped absorption features appear in all velocity ranges, indicating that they originate from some continuum absorption mechanism.
In addition, we find regions of weaker and stronger line-shaped absorption features spatially connected to each other appearing as dark and bright line-shaped features around knots A, B, and C. This has not been seen in any previous studies in which the M87 jet was usually seen in emission only. 

\begin{figure*}[ht!]
\centering
\includegraphics[width= 11cm, height = 7cm]{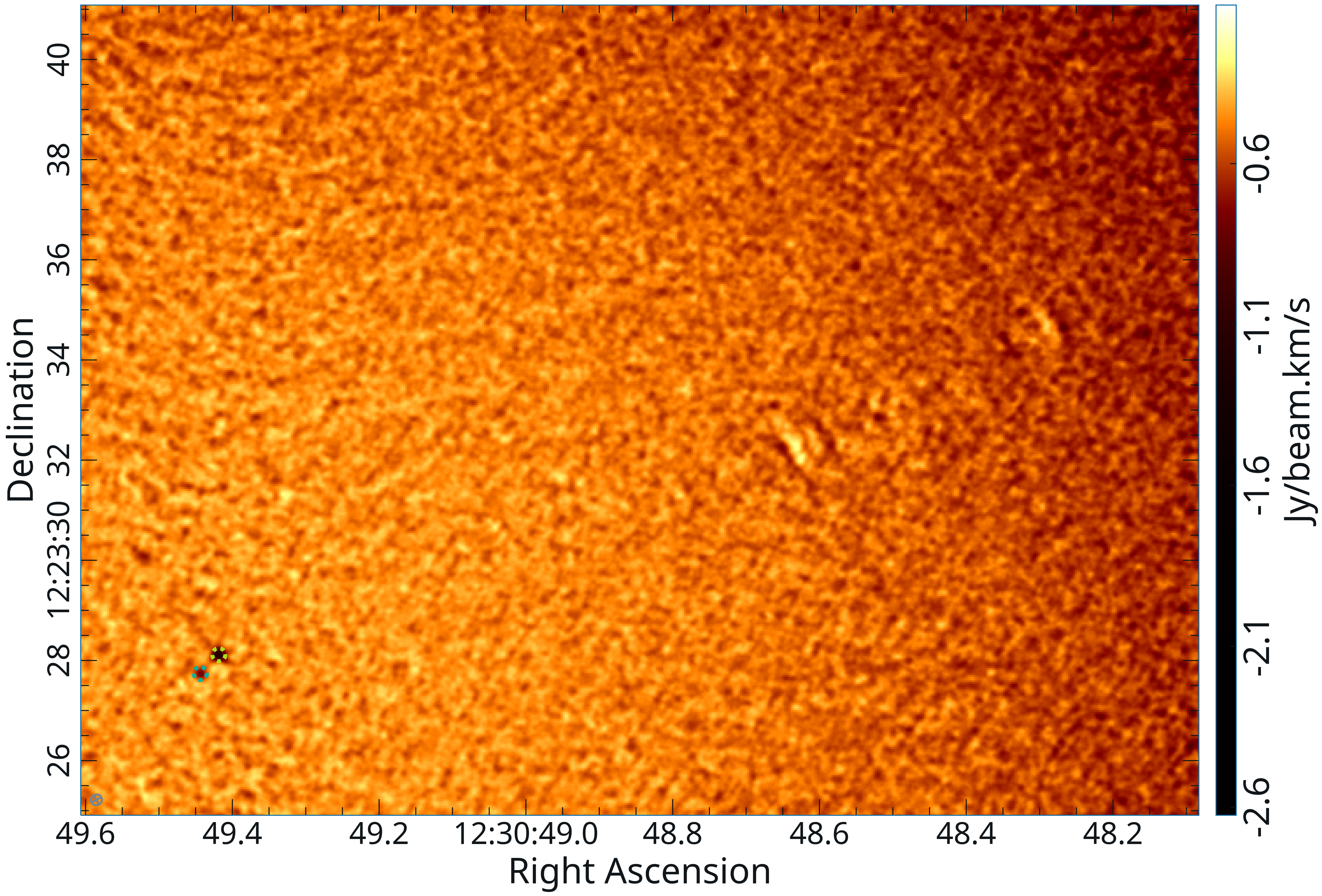}
\includegraphics[width= 11cm, height = 7cm]{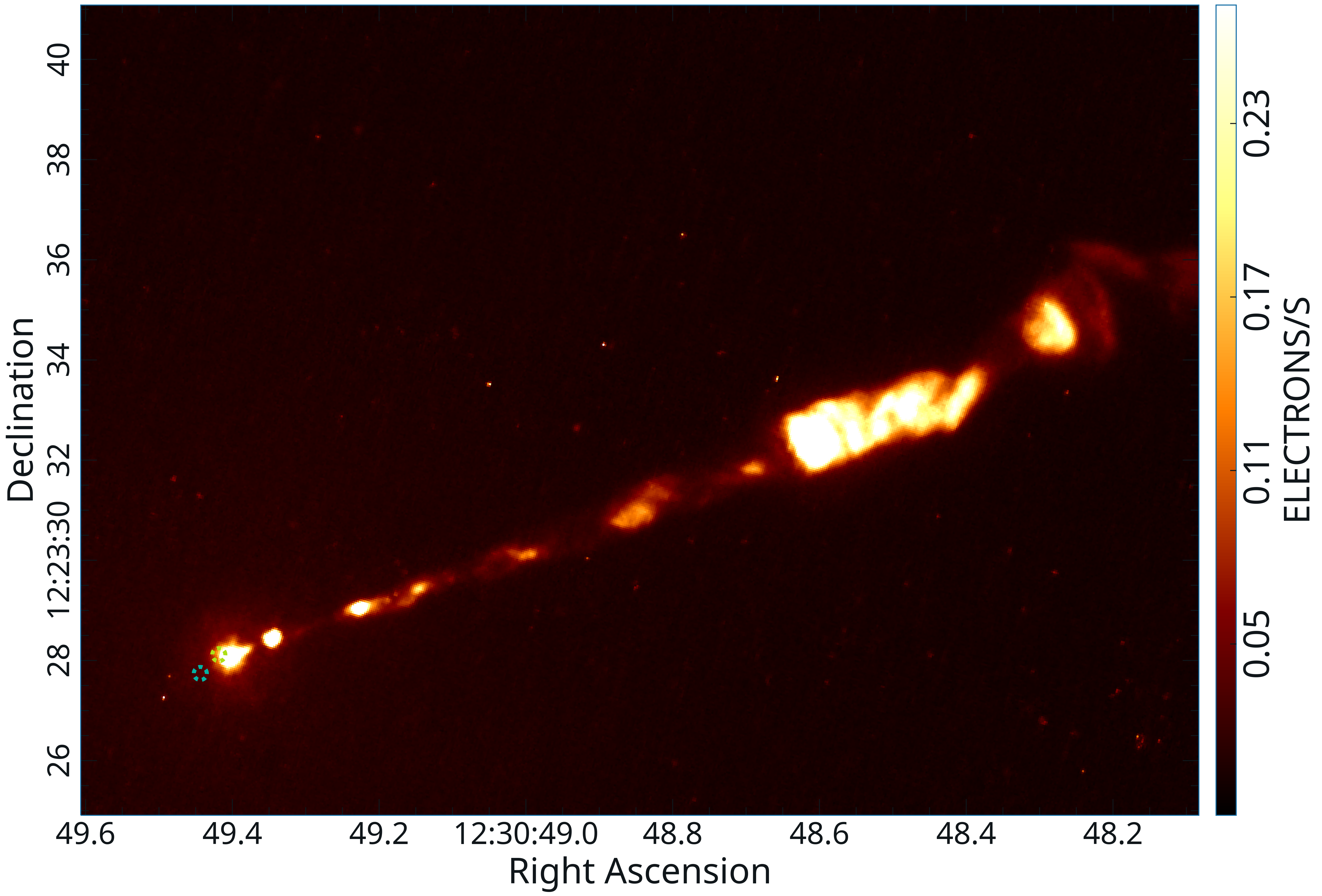}
\includegraphics[width= 11cm, height = 7cm]{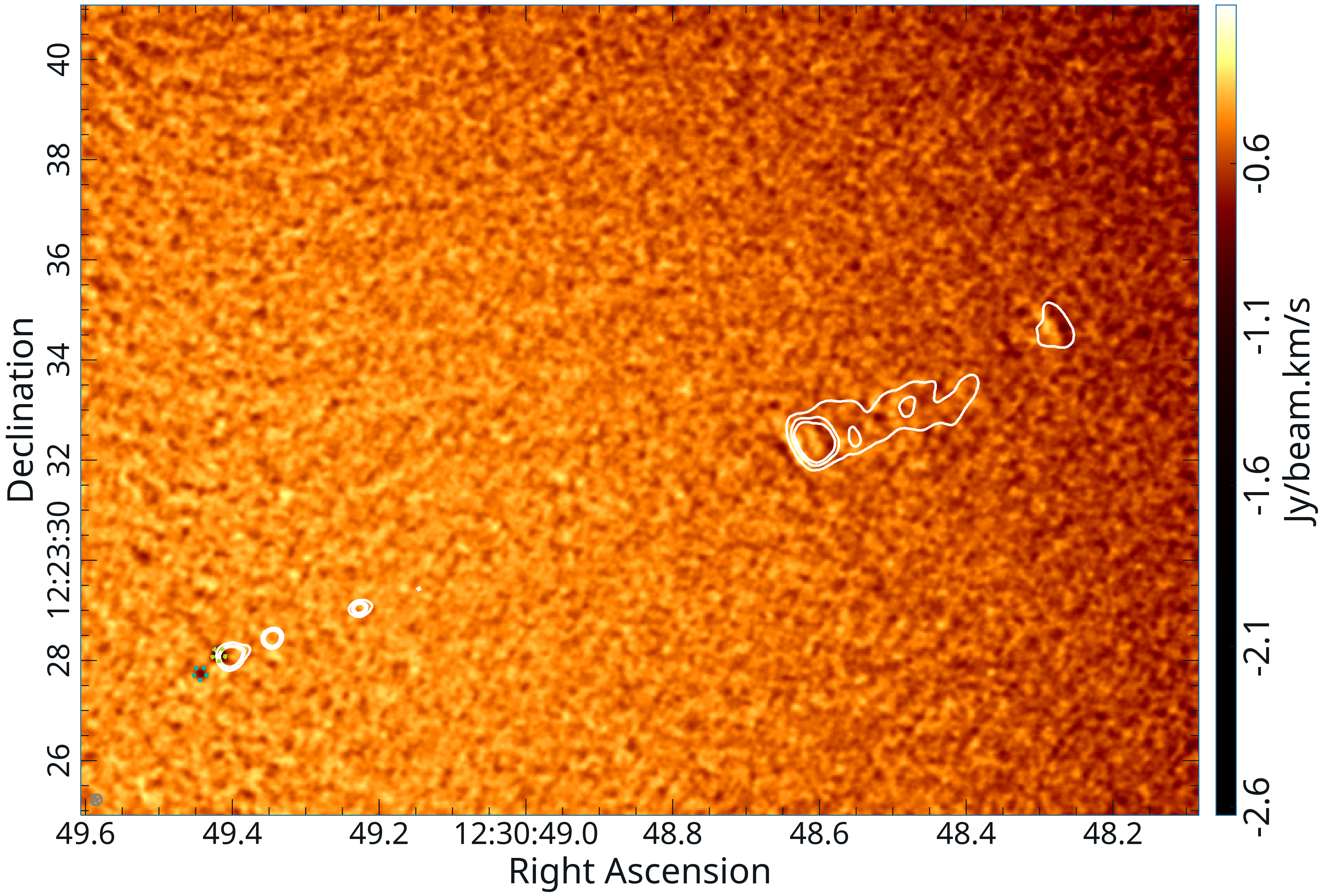}
\caption{Comparison of the ALMA integrated absorption map (top) with the HST data (middle) (reference: ads/Sa.HST\#IC0601030), the green dashed circle shows the location of ALMA-1 and the sky blue dashed circle shows ALMA-2. Bottom image shows our ALMA data with the HST contours to show the location matching of the shock observed regions. White contours are drawn at 3, 6 and 9 $\sigma$ of HST image of the M87 nucleus and jet.}
\label{fig:ALMA and HST comparison}
\end{figure*}


\subsection{Interpretations for the origin of the compact features} \label{subsec:origin of the features}
To investigate the origin of the compact absorption features and constrain the nature of these
objects, we extracted the individual spectra of ALMA-1 and ALMA-2 with a selected size of $0.26\arcsec \times 0.26\arcsec$. The resulting profiles are shown in Figure 5. The spectral features of ALMA-1 and ALMA-2 eliminate several possibilities. Because it appears only within a certain velocity range, it cannot originate from continuum absorption. Its absence in a broader velocity range rules out artifacts associated with continuum side-lobes or CLEAN residuals. Moreover, it cannot be caused by problems in bandpass calibration because these two compact features appear only in specific spatial regions.

\begin{figure*}[ht!]
\centering
\includegraphics[width=11cm, height=7cm]{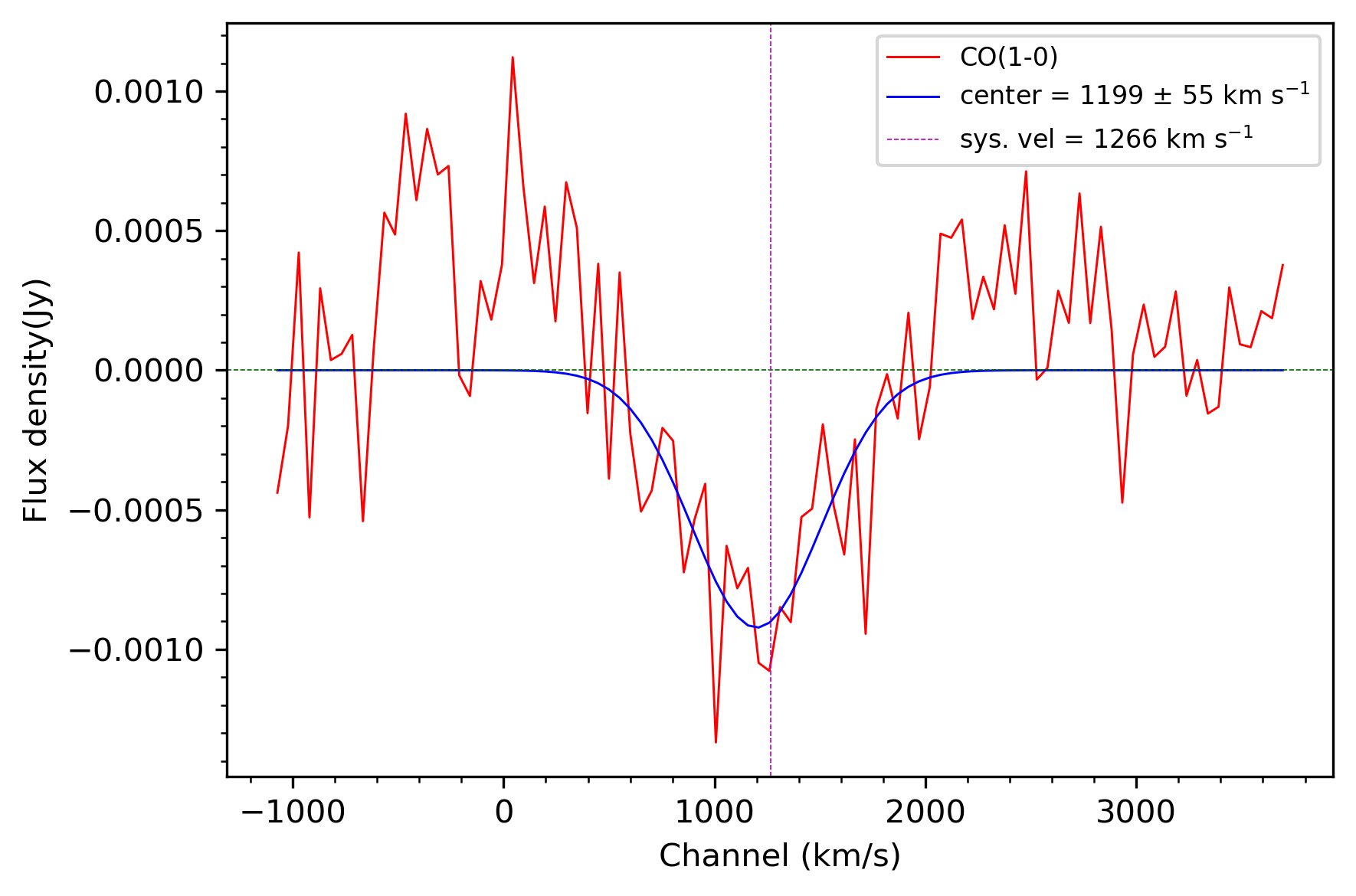}
\includegraphics[width=11cm, height=7cm]{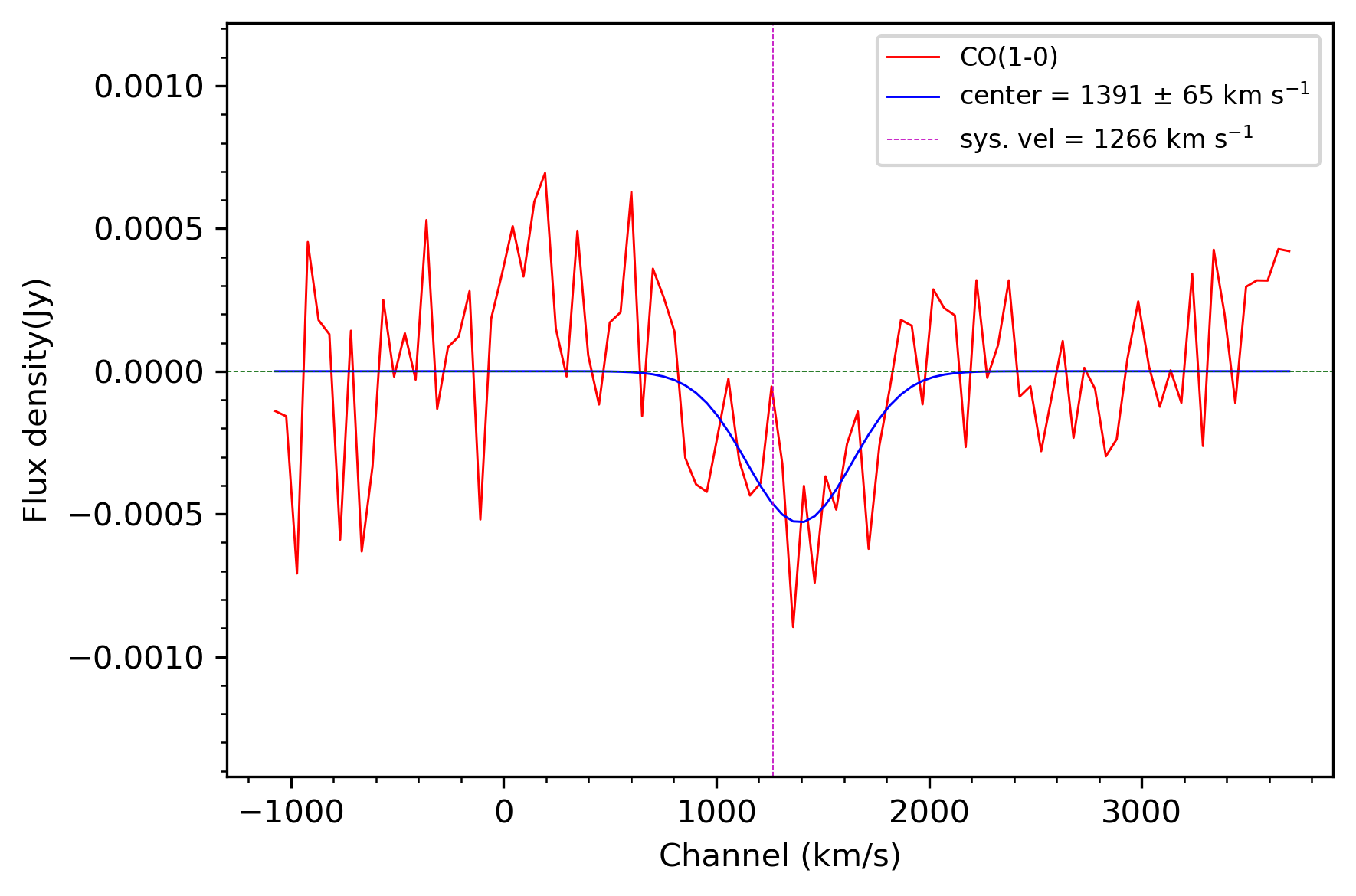}
\caption{Observed spectra of the ALMA-1 and ALMA-2 regions. The blue curve shows the best-fit Gaussian model.}.
\label{fig:ALMA-1 and ALMA-2 spectra}
\end{figure*}

A possible absorption mechanism is caused by CO(1-0). A single giant molecular cloud (GMC) is ruled out because the observed absorption velocity dispersion is much broader than the velocity distribution of a single GMC. Therefore, plausible interpretations would require many molecular clouds or filaments over a huge velocity range. A similar structure can be found in AGNs, i.e. molecular torii or molecular clumps surrounding SMBHs. In addition, we found the absorption profile in the spectrum to be similar to the results in \citet{Ray...} which was observed in much lower resolution and was interpreted as the molecular clouds surrounding the SMBH of M87.

In this scenario, the compact absorption that appears at the position of ALMA-1 is a natural outcome of the absorption of molecular clouds surrounding the SMBH of M87. However, ALMA-2 is $\sim 40~\mathrm{pc}$ away from ALMA-1 and should have a different origin.

One possibility is to interpret ALMA-2 as an absorption feature caused by a secondary SMBH. Previous studies suggested the possibility of a secondary SMBH ($4\times10^7 -3\times10^8~\mathrm{M_\odot}$) at a separation of $\sim 0.01-0.05~ \mathrm{pc}$ from the primary one \citep{secondary_BH...} and/or a massive object $\mathrm{\sim7~pc}$ away from the central SMBH in the counter-jet direction \citep{SMBH_offset...}. In the integrated absorption map, we identify a compact absorption feature spatially offset by $\sim\mathrm{0.5\arcsec~(\sim 40~pc)}$ in addition to the absorption feature that coincides with the central SMBH. It raises the possibility that ALMA-2 traces molecular gas associated with a secondary SMBH. Therefore, here we suggest a possible secondary SMBH at a distance of $\sim 40~\mathrm{pc}$ from the primary SMBH of M87.  

A secondary SMBH at a projected distance of $\sim 40$ pc would have an orbital time scale of the order $\sim 10^5$ yr and would not be expected to produce observable short-timescale signatures. From the spectrum of ALMA-1 and ALMA-2 we find a relative velocity difference of $\sim \mathrm{192~km~s^{-1}}$, whereas a typical orbital velocity should be approximately $\mathrm{\sim 1000~km~s^{-1}}$. However, the actual separated distance should be larger than the projected separation of $\sim 40$ pc and the true orbital velocity should be lower. Besides, the relative velocity was observed in the radial direction and represents only the projection of the true velocity.
The secondary SMBH only shows weak evidence of AGN activity and seems to be quiet. This could be caused by the fact that the second SMBH had been stripped of the gas and material that need to power the AGN when it fell into the Virgo galaxy cluster.  

The mass of the secondary SMBH is difficult to estimate. If we assumed that the two SMBH are bound pair and they are rotating at the center of the mass of the system, which is at a distance of $\sim 7$ pc towards the counter-jet direction from ALMA-1 as suggested by \citet{SMBH_offset...}, we obtained the mass of the second SMBH $\mathrm{M_2 \approx 1.3 \times10^{9}\,M_{\odot}}$ assuming the mass of the primary SMBH $\mathrm{M_1 = 6.5 \times10^{9}\,M_{\odot}}$ \citep{EHT...} .
We can also use the velocity dispersion ratio of ALMA-1 and ALMA-2 to estimate the mass of the secondary SMBH since $\mathrm{M \propto R\sigma^2}$, where R is the size of the distribution region of the gas. We assume that the size is proportional to the peak absorption of the gas; from the spectral profiles of ALMA-1 and ALMA-2 we obtained the mass of the ALMA-2 is $\mathrm{\approx 2.2 \times10^{9}\,M_{\odot}}$, assuming the mass of the primary SMBH is $\mathrm{ 6.5 \times10^{9}\,M_{\odot}}$ \citep{EHT...}. 

\subsection{Interpretation for the line-shaped absorption features}

The line-shaped absorption features are caused by continuum absorption. Since the jet regions are ionized regions, the absorption features are likely to be caused by electron scattering of the background photons. The stronger absorption features are denser regions of electrons and vice versa. The detected spatially connected dark and bright line-shaped features suggest that existence of spatially connected high-density and low-density regions of electrons. We suggest that the high and low density regions are produced by shocks between the jet and the ambient ISM. 

\section{Conclusion} \label{sec:conclusion}

Our analysis of ALMA Band~3 observation reveals two compact 
absorbers, one located directly along the line-of-sight to the M87 AGN and the other $\mathrm{\sim 40~pc}$ away from the AGN. The features are detected only within a velocity range 
($-500$ to $+2000~{\rm km~s^{-1}}$).

We interpret these two absorbers as cold, dense molecular fragments embedded within the inner few parsecs of AGNs. Therefore, we interpret that the ALMA-2 feature arises from the presence of a secondary SMBH in M87. Our results provided a new evidence for the existence of a secondary SMBH in the M87 system.

We also found line-shaped absorption features perpendicular to the M87 jet. The line-shaped features are caused by some continuum absorption mechanisms. We interpreted this absorption was caused by electron scattering of background photons and the line-shaped features are high and low density regions caused by shocks between the jet and the ambient ISM. 

\begin{acknowledgments}
This paper makes use of the following ALMA data:  ADS/JAO.ALMA-2016.1.00021.S ALMA is a partnership of ESO (representing its member states), NSF (USA) and NINS (Japan), together with NRC (Canada), NSTC and ASIAA (Taiwan), and KASI (Republic of Korea), in cooperation with the Republic of Chile. This project was supported by Ministry of science and technology (MOST) or National Science and Technology council (NSTC), grant no NSTC 113-2112-M-008-001 and NSTC 114-2112-M-008-018. In addition, I would like to thank the Institute of Astronomy and astrophysics for funding the publication of this paper.
\end{acknowledgments}

\begin{contribution}

All authors contributed equally to this paper.


\end{contribution}

\bibliography{reference}{}
\bibliographystyle{aasjournalv7}



\end{document}